%% file: main.tex
\documentclass[preprint,12pt]{elsarticle}
\usepackage[colorinlistoftodos]{todonotes}
\usepackage{csquotes}
\usepackage{longtable}
\usepackage{graphicx}
\usepackage{adjustbox}
\usepackage{float}
\usepackage[inline]{enumitem}
\usepackage{color, colortbl}
\usepackage{svg}
\usepackage{amsmath}
\usepackage{paralist}
\usepackage{enumitem}
\usepackage{pdfpages}
\usepackage{multirow}
\usepackage{booktabs}

\newlist{questions}{enumerate}{4}
\setlist[questions,1]{label=RQ\arabic*.,ref=RQ\arabic*}
\setlist[questions,2]{label=(\alph*),ref=\thequestionsi(\alph*)}
\setlist[questions,3]{label=(\alph*),ref=\thequestionsi(\alph*)}
\setlist[questions,4]{label=(\alph*),ref=\thequestionsi(\alph*)}

\newlist{inclusion}{enumerate}{2}
\setlist[inclusion,1]{label=IC\arabic*.,ref=IC\arabic*}
\setlist[inclusion,2]{label=(\alph*),ref=\theinclusioni(\alph*)}

\newlist{exclusion}{enumerate}{9}
\setlist[exclusion,1]{label=EX\arabic*.,ref=EX\arabic*}
\setlist[exclusion,2]{label=(\alph*),ref=\theexclusioni(\alph*)}
\setlist[exclusion,3]{label=(\alph*),ref=\theexclusioni(\alph*)}
\setlist[exclusion,4]{label=(\alph*),ref=\theexclusioni(\alph*)}
\setlist[exclusion,5]{label=(\alph*),ref=\theexclusioni(\alph*)}
\setlist[exclusion,6]{label=(\alph*),ref=\theexclusioni(\alph*)}
\setlist[exclusion,7]{label=(\alph*),ref=\theexclusioni(\alph*)}
\setlist[exclusion,8]{label=(\alph*),ref=\theexclusioni(\alph*)}
\setlist[exclusion,9]{label=(\alph*),ref=\theexclusioni(\alph*)}

\newlist{bknowledge}{enumerate}{5}
\setlist[bknowledge,1]{label=\arabic*.,ref=\arabic*}
\setlist[bknowledge,2]{label=(\alph*),ref=\thebknowledgei(\alph*)}
\setlist[bknowledge,3]{label=(\alph*),ref=\thebknowledgei(\alph*)}
\setlist[bknowledge,4]{label=(\alph*),ref=\thebknowledgei(\alph*)}
\setlist[bknowledge,5]{label=(\alph*),ref=\thebknowledgei(\alph*)}

\usepackage{amssymb}

\usepackage{lineno}
\usepackage{multirow}

\usepackage{booktabs}

\journal{Information and Software Technology}

\begin{document}

\begin{frontmatter}




\title{Synthesizing Research on Programmers' Mental Models of Programs, Tasks and Concepts -- a Systematic Literature Review} 

\author[myu]{Ava Heinonen \corref{cor1}}
\ead{ava.heinonen@aalto.fi}

\author[myu]{Bettina Lehtel\"a}
\author[myu]{Arto Hellas}
\author[myu]{Fabian Fagerholm}

\address[myu]{Department of Computer Science, Aalto University, P.O.Box 15400, FI-00076 AALTO, Helsinki, Finland}

\cortext[cor1]{Corresponding author}


\begin{abstract}
  \input{00-abstract}

\end{abstract}


\end{frontmatter}



\newpage
\section{Introduction}
\label{sec:introduction}

\input{10-introduction}

\section{Background}
\label{sec:background}
\input{20-background}

\section{Research Method}
\label{sec:methodology}
\input{30-methodology}

\section{How Have Mental Models Been Studied?}
\label{sec:res_rq1}
\input{41-results_RQ1}

\section{How Have Programmers’ Mental Models Been Conceptualized and Described?}
\label{sec:res_rq2}
\input{42-results_RQ2}

\section{Synthesis of Existing Results}
\label{sec:res_rq34}
\input{43-results_RQ3-4}

\section{Discussion}
\label{sec:discussion}
\input{50-discussion}

\section{Conclusions}
\label{sec:conclusions}
\input{60-conclusions}


\appendix
\section{Data Extraction Form}
\label{ap:form}
\section{Items Included in The Result Set}
\label{ap:items}





\bibliographystyle{plain}
\bibliography{references}

\end{document}

%% file: 00-abstract.tex
\emph{Context}: Programmers’ mental models represent their knowledge and understanding of programs, programming concepts, and programming in general. They guide programmers’ work and influence their task performance.  Understanding mental models is important for designing work systems and practices that support programmers. 

\emph{Objective}: Although the importance of programmers’ mental models is widely acknowledged, research on mental models has decreased over the years. The results are scattered and do not take into account recent developments in software engineering. In this article, we analyze the state of research into programmers’ mental models and provide an overview of existing research. We connect results on mental models from different
strands of research to form a more unified knowledge base on the topic.  

\emph{Method}: We conducted a systematic literature review on programmers' mental models. We analyzed literature addressing mental models in different contexts, including mental models of programs, programming tasks, and programming concepts.  Using nine search engines, we found 3678 articles (excluding duplicates). Of these, 84 were selected for further analysis. Using the snowballing technique, starting from these 84, we obtained a final result set containing 187 articles.

\emph{Results}: We show that the literature shares a kernel of shared understanding of mental models. By collating and connecting results on mental models from different fields of research, we uncovered some well-researched aspects, which we argue are fundamental characteristics of programmers' mental models.

\emph{Conclusion}: This work provides a basis for future work on mental models. The research field on programmers' mental models still faces many challenges rising from a lack of a shared knowledge base and poorly defined constructs.  We created a unified knowledge base on the topic.  We also point to directions for future studies.  In particular, we call for studies that examine programmers working with modern practices and tools.

%% file: 10-introduction.tex
The daily tasks of programmers -- developing, maintaining and debugging programs -- require a comprehensive understanding of programming languages and frameworks, programming concepts, and details of the given programming task~\cite{robertson1990knowledge, davies1993models, bisant1993cognitive}. Due to the importance of knowledge in programmers’ work, understanding it plays an important role in designing tools and systems that support programmers’ task performance. Therefore, gaining insight into the aspects of human cognition and knowledge that underpin programmers’ task
performance is of great value to the field of software engineering~\cite{fagerholm2021}.

In this article, we describe and discuss programmers' understanding as \emph{mental models}. A mental model refers to the person's internal representation of a target system. It is their internal model of what something is and how it works~\cite{van2021reflections}. In programming contexts, mental models represent programmers' knowledge and understanding of code~\cite{burgos2007through, yates2020characterizing}, programming concepts~\cite{sajaniemi2005investigation},  programming tasks~\cite{balijepally2012effect}, and programs being designed and developed~\cite{dawson2013cognitive}.  

It is difficult to estimate whether existing research can describe and explain programmers' mental models in contemporary settings.  
The term mental model is used in research into program comprehension~\cite{von1995industrial}, computing education~\cite{sajaniemi2008study},  software engineering practices~\cite{yates2020characterizing} and tool development~\cite{kang2017omnicode}. But the research field has not yet reached agreement on what it means to understand a program, a concept, or a task \cite{harth2017program}. Furthermore, the research field has yet to build a shared knowledge base and a unified conceptualization of mental models. This has led to a lack of shared language \cite{detienne2001software} and scattered research efforts~\cite{bidlake2020systematic}.  Therefore, many studies do not build on existing results~\cite{bidlake2020systematic, bisant1993cognitive, von1995program}.  The fragmentation of the knowledge base has resulted in a situation where existing results are difficult to connect into a coherent whole~\cite{bidlake2020systematic}. 

Our study addresses the following research questions:
\begin{questions}
\item How have programmers' mental models been studied?
\item How have programmers' mental models been conceptualized and described?
\item What are the central aspects of programmers' mental models present in the literature? 
\item What can be concluded about programmers' mental models based on the results related to these central aspects?
\end{questions}
To answer these questions, we performed a systematic literature review (SLR) following well-recognized guidelines~\cite{keele2007guidelines}.  

This article is organized as follows. In Section~\ref{sec:background}, we describe the background of the study. In Section~\ref{sec:methodology}, we detail how our systematic literature review was conducted. In Sections~\ref{sec:res_rq1}, \ref{sec:res_rq2}, and \ref{sec:res_rq34}, we present the results of the review. In Section~\ref{sec:discussion}, we further discuss the results in relation to our research questions. We also discuss the validity of the study. Finally, in Section~\ref{sec:conclusions}, we present concluding remarks and provide directions for future research.

%% file: 20-background.tex
In this section, we describe the background literature relevant to our study and discuss related reviews in this field. 

\subsection{Mental Models}
Mental models are conceptualizations for the way a person perceives a target system. These target systems can be physical and abstract systems, abstract knowledge domains, or situations~\cite{jones2011mental,staggers1993mental, rickheit19991}. Mental models represent an understanding of what the target system is. This includes its behavior and how its parts fit together~\cite{van2021reflections}.  These models allow people to reason about, predict, and explain the behavior of the target system~\cite{van2021reflections}. 
The idea of mental models can be applied to understanding how people interact and use different physical and software systems~\cite{jones2011mental,staggers1993mental}. Thus, various studies in computer science have used mental models to explain programmers' understanding of programs~\cite{van2021reflections, rickheit19991}, programming-related tasks~\cite{balijepally2012effect,balijepally2015task} and concepts~\cite{sajaniemi2005investigation}.

Understanding programmers' mental models has a direct impact on understanding their work. It is also important for designing tools and practices to support it. Developers spend more than half of their time comprehending code~\cite{xia2017measuring}. This leads to a need for tools and methods to support this activity~\cite{maalej2014comprehension}.  In recent years, knowledge of programmers' mental models has been utilized in different practical applications~\cite{henley2018codedeviant,kang2017omnicode,bradley2018context}. However, these tools do not find their way into the daily use of developers~\cite{maalej2014comprehension}. It has been proposed that this is not due to these tools being not needed. Rather, it is due to the fact that our current understanding of programmers' mental models is not detailed enough to develop beneficial tools \cite{harth2017program}. 

Multiple theories have been presented to analyze how programmers acquire mental models of programs and how knowledge is represented in their minds~\cite{detienne2002understanding, von1995industrial}. Existing theories from text comprehension and schema theory have also been adapted to explain program comprehension and programmers' mental models~\cite{detienne1990program}. Although the most popular theories share some common core features, they differ in their level of explanation and details of mental models and the acquisition process~\cite{von1995program}. 

Text-comprehension-based theories treat code comprehension as reading text~\cite{detienne2002understanding}.  According to these theories, programmers' understanding of the code syntax and its semantic meaning are linked, but separate, mental models~\cite{burkhardt2002object, alardawi2015novice, burkhardt1997mental, corritore1999mental, khazaei2002there, wiedenbeck1999novice}. Theories differ in their understanding of the program comprehension process~\cite{von1995program}.  Some theories propose that the process is bottom-up -- it starts from understanding the code syntax, allowing the programmer to then derive the semantic meaning of the code from the syntax~\cite{von1995program}. Some theories suggest that the process is top-down, starting from a higher-level hypothesis of code functionality, which is then verified and defined by investigating the code syntax~\cite{detienne2002understanding}.

Schema theory has also been used to explain the mental models of programmers ~\cite{detienne1990program, romero2001focal, balijepally2012effect}. This approach is based on programmers' background knowledge of reoccurring code patterns they accumulate through experience and instruction--schemata. According to schema theory, program comprehension is driven by schema activation, or understanding the meaning of code syntax through matching it with existing background knowledge. The mental model can then be described as a collection of these activated schemata~\cite{romero2001focal, balijepally2012effect}. Different comprehension theories have also been integrated into the so-called integrated metamodel theory \cite{von1995program}. According to the integrated metamodel theory, program comprehension uses top-down and bottom-up processes to seek information on the semantic and syntactic aspects of the program~\cite{von1994comprehension}. It also states that the comprehension process relies on  programmers' background knowledge and information in the program itself \cite{von1994comprehension}.
 
Mental models of programming concepts have been studied mainly in the context of computer science education. Students' mental models of programming-related concepts have been the topic of interest in many studies. These mental models represent the students' understanding of programming concepts~\cite{gotschi2003mental, pieterse2018triviality, ma2007investigating}.  Studies have classified students' mental models of concepts such as recursion~\cite{bhuiyan1989mental, bhuiyan1991characterizing}. This provides educators with information on how these concepts are understood and leads to research-informed ways of teaching them.

 Task mental models are seen to contain an understanding of the task and an understanding of the solution.  In the case of program modification, they also contain an understanding of the program to be modified \cite{balijepally2012effect, bradley2018context, shaft2006role}. In some studies, a task mental model refers to the programmer's understanding of all three: the program that is being modified, the required modifications and the changes required to modify the program \cite{balijepally2012effect}. In other studies, the program being modified, the task context, is understood to be a separate mental model \cite{bradley2018context}, while the programmer's understanding of the task represents another \cite{shaft2006role}.

\subsection{Related Reviews}
 Researchers have examined the concept of mental models and evaluated the research field in previous publications. 

Staggers and Norcio~\cite{staggers1993mental} explored mental models in human-computer interaction. They analyzed existing theories of mental models and provided an explanation of what mental models mean in the context of human-computer interaction. However, the review is thirty years old and the research field has evolved. In this study, we have included the last thirty years of research to update the picture of mental models.  

A recent systematic literature review included 72 studies~\cite{bidlake2020systematic}. The study evaluated research on programmers' mental representations of code in programming situations. In this study, we complement their research by adding research on mental models of tasks and concepts. 

Another recent review provided a synthesis of existing code comprehension theories~\cite{fekete2020comprehensive}. This review examined different cognitive models of program comprehension. The focus of the study was on the acquisition of mental models.  Our study complements their research by adding mental model structure to the analysis. 

Studies in other disciplines have also analyzed mental models. For example, Rook~\cite{rook2013mental} assessed the concept of mental models to build a definition of mental models for organizational management. Jones et al.~\cite{jones2011mental} completed a similar study analyzing research on mental models from multiple disciplines to provide the field of natural resource management with a unified model of mental models. 

%% file: 30-methodology.tex
In this section, we describe the method used to conduct the systematic literature review.  Figure~\ref{fig:process} shows an overview of the process.

\subsection{Search Strategy}
Our search strategy consisted of two stages: the pilot search and the search stage. 

\subsubsection{Pilot Search}
\label{sec:pilot-search}
During the pilot search stage, we created a pilot set of articles that was used to create the search string and select article databases. We selected three articles on programmers' mental models~\cite{hoc1977role,hoc1990language,brooks1983towards} and a recent systematic literature review~\cite{bidlake2020systematic} as starting points. We did a step of forward and backward snowballing, which produced 75 articles. We then performed manual searches using the keywords identified so far in the process, which yielded 22 articles. The results were evaluated on the basis of their title and abstract. After evaluation, the pilot set contained 52 articles. 

\subsubsection{Search String}
\label{sec:database-search}
 We extracted relevant terms and synonyms for target populations, mental models, and target systems from items in the pilot set. To form the search string, we combined terms for (1) the target population, (2) mental models, and  (3)relevant target systems. We tested multiple iterations of the search string using different terms and synonyms. We evaluated the number of results, the number of relevant results, and the degree to which searches were able to find articles in the pilot set. The chosen search string captured most relevant articles while excluding most irrelevant results. 

We used two versions of the search string. In databases that support the W/$n$ operator, or the synonymous NEAR/$n$ operator, we used the following search string: 
\begin{verbatim}
    1. ( ( programmer OR coder OR developer )
    2.   AND ( mental W/2 ( model OR representation ) )
    3.   AND ( code OR program* OR software OR comput* ) )
\end{verbatim}
In databases that did not support these operators, we used the following search string:
\begin{verbatim}
    1. ( ( programmer OR coder OR developer )
    2.   AND ("mental model" OR "mental models" OR
         "mental representation" OR "mental representations")
    3.   AND ( code OR program* OR software OR comput* ) )
\end{verbatim}

 In databases that supported limiting the search to title, abstract, and keywords, this filtering was used. In databases that did not allow for this filtering, the search was performed on the fields available in those databases. 

\subsubsection{Database Selection}
We used the items in the pilot set to identify and select relevant databases. The nine selected databases and the number of results from each database are described in Table~\ref{tab:dbtable}. We performed the queries on 30 June and 1 July 2020.

\renewcommand{\arraystretch}{1.2}
\definecolor{Gray}{gray}{0.9}
\begin{table}[H]
    \centering
    \begin{tabular}{l l}
    \textbf{Database} & \textbf{Number of results} \\
    \toprule
    Scopus & 210 \\
    Science Direct & 133 \\
    ACM &  369 \\
    Springer &  3220 \\
    Taylor and Francis &  2 \\
    Web of Science &  141 \\
    IEEE Xplore &  70 \\
    Sage &  45 \\
    Wiley & 157 \\
    \addlinespace
    \rowcolor{Gray}
    Total &  4283 \\
    \bottomrule
    \end{tabular}
    \label{tab:dbtable}
    \caption{Databases used in the search and the number of results from each database.}
\end{table}

\subsection{Inclusion and Exclusion Criteria}
\label{sec:inclusion-exclusion-criteria}
We defined two inclusion criteria:

\begin{inclusion}
\item Studies whose main focus is on studying the structure or acquisition of programmers' or programming students' mental models of different target systems or programming activities. 
\item Studies whose main focus is on studying the effects of mental models on the use of different target systems or the execution of different software development activities. 
\end{inclusion}

We also defined nine exclusion criteria:

\begin{exclusion}
    \item Papers on shared mental models or group mental models.
    \item Papers on specific brain patterns or brain activity when no link to mental models is demonstrated.
    \item Papers on technological interventions or tools developed to display, describe, or communicate mental models if the article does not contain research and results related to mental models themselves.
    \item Papers concerning the development of artificial mental models for artificial intelligence or robots.
    \item Papers concerning users' mental models or end-user development except for developers or software development students using an end-user-programming system for software development activities. 
    \item Papers on non-developers' mental models of target systems or development activities. 
    \item Papers not written in English.
    \item Short papers, such as extended abstracts or idea papers. 
    \item Non-academic sources where the quality of the information cannot be determined. 
\end{exclusion}

\subsection{Study Filtering Process}
\label{sec:filtering}
The set of results acquired from database searches was filtered to exclude irrelevant results. 
\subsubsection{Duplicate Removal and Initial Filtering}
First, all results were combined to form the initial result set. An automated duplicate check was performed using the Mendeley Desktop duplicate check tool. Duplicates were verified and merged using the Mendeley Desktop tools. This removed 605 duplicate results, leaving 3678 results. 

Duplicate removal was followed by initial filtering. During the initial filtering, items in the initial result set were evaluated based on their type, language, and title. This was performed by one researcher and removed 2916 items, leaving 762 items in the the initial result set.

\subsubsection{Exclusion Rounds}
Initial filtering was followed by two exclusion rounds, E1 and E2. During E1 and E2, the initial result set was divided between the research team to evaluate and decide on inclusion or exclusion. Another researcher verified all exclusion decisions. If the researchers disagreed, they discussed the decision to reach agreement, consulting the other researchers if necessary. During E1, items were evaluated based on their title. During E2, they were evaluated based on title and abstract. The two exclusion rounds excluded 566 items, leaving 196 items in the initial result set.

\begin{figure}[H]
  \centering
  \includegraphics[width=\textwidth]{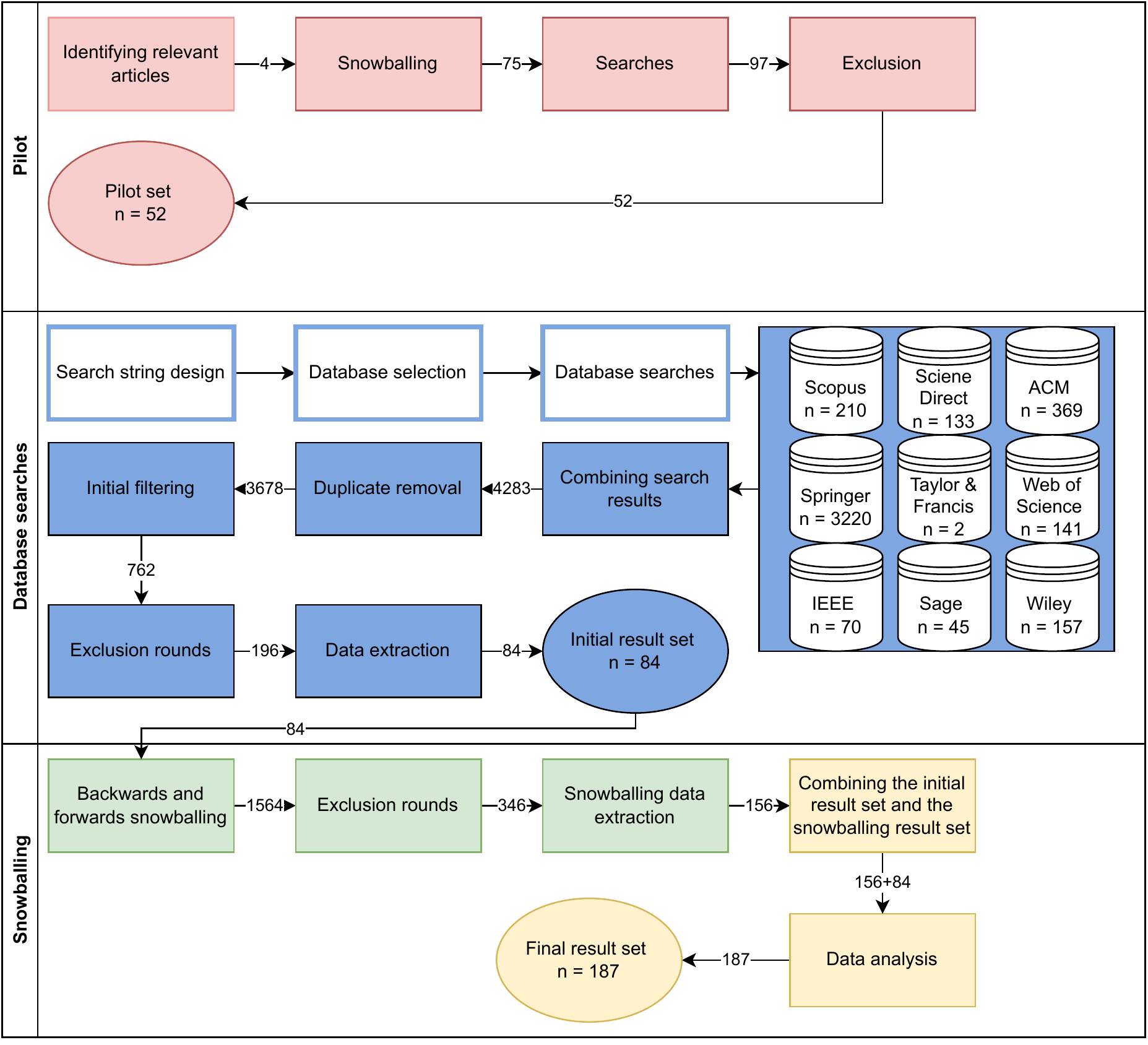}
  \caption{Filtering process and the size of the pilot set, the initial result set, the snowballing set, and the final result set at each point in the process. \label{fig:process}}
\end{figure}

During data extraction, another 112 items were excluded, leaving 84 items in the initial result set.  Figure~\ref{fig:process} shows the study filtering process and details the number of items included in each step of the process.

\subsection{Snowballing}
\label{sec:snowballing}
To achieve better coverage, we performed one round of forward and backward snowballing, following Mourao et al.~\cite{mourao2020performance}.
In backward snowballing, the reference list of each item in the initial result set was evaluated and relevant references were included. In forward snowballing, Scopus and Google Scholar were used to find publications that cite each item in the the initial result set. Publications were evaluated based on their title and abstract, and the relevant publications were included.
After removing duplicate results, we had 1564 items in the snowballing set. It was then filtered using the process described in~\ref{sec:filtering}. After filtering, the snowballing set contained 346 items. During data extraction, further 190 items were excluded. The final result set contains both the initial result set and the snowballing result set. Further 53 items were excluded during data analysis. The final result set therefore contained 187 results. 

\subsection{Data Extraction}
We designed a data extraction form to extract data from the results in a systematic way. A summary of the data extraction form is provided in Table \ref{tab:dataextractionform}.
For details of the data extraction form, see the supplementary material in~\ref{ap:form}.

\definecolor{Gray}{gray}{0.9}

\begin{longtable}{p{55mm} p{50mm}}
\textbf{Field} & \textbf{Categories} \\
\toprule
\rowcolor{Gray}
\multicolumn{2}{l}{\emph{Participants and Experience}} \\
\textbf{Experience Level} & Novice \\
& Intermediate \\ 
& Expert \\
& Varying \\
& NOS  \\
& Other \\
\addlinespace
\textbf{Participant Type} & Professional \\
& Student \\
& PhD student  \\
 & Academic  \\
\textbf{Experience type} & General \\
& Task specific\\ 
\textbf{Experience measure} & Self-reported \\
& Course performance \\
& Task performance \\
& Level of education \\
& Time in employment \\
& Employment in the field \\
\rowcolor{Gray}
\multicolumn{2}{l}{\emph{Target System}} \\
\textbf{Target System}  & Task\\
& Program  \\
& Concept \\
& Paradigm \\
& NOS  \\
& Other \\
\textbf{Program Type} & Application \\
& Fragment \\
& Not specified \\
\rowcolor{Gray}
\multicolumn{2}{l}{\emph{Study Details}} \\
\textbf{Study setting} & Natural \\
& Natural  \\ 
& Laboratory  \\ 
& Online\\
\textbf{Study format} & Coding task \\
& Other task \\
&Interview  \\
\rowcolor{Gray}
\multicolumn{2}{l}{Theoretical concepts} \\
\textbf{Structure Concepts} & S.R.T \\
& Program model  \\
& Situational model   \\
& Domain model  \\
& Schema \\
& Programming plans \\
& Text structure knowledge \\
& Rules of discourse \\
& Chunk \\
\textbf{Acquisition Concepts} & Bottom-up \\
& Bottom-up \\
& Top-down \\
& Combination \\
& Systematic \\
& As-needed \\
& Hypothesis-based \\
& Expectation-based  \\
& Inference-based  \\
& Chunking \\
& Beacons \\
& Interaction with the target system \\
& Knowledge-based \\
\rowcolor{Gray}
\multicolumn{2}{l}{\emph{Mental Model Descriptions}} \\
\textbf{Mental Model Description} & How MM were described and conceptualized \\
\rowcolor{Gray}
\multicolumn{2}{l}{\emph{Results}} \\
\textbf{Results} & MM Structure \\
& MM Acquisition  \\
& Effects of MM in Task Performance  \\
& Experience \\
& Context \\
\bottomrule
\caption{Summary of the data extraction form, detailing the fields of the form used in the analysis in this article. MM = Mental Model}
\label{tab:dataextractionform}
\end{longtable}

The data extraction form was created by the first author. An evaluation meeting was held with all authors to assess the form. The form was further evaluated and adjusted before snowballing after data was extracted from items in the initial result set. 

To calibrate consensus between researchers, a series of trial extractions was performed. All authors extracted data from the same articles. A series of meetings was held to discuss extraction. These meetings were held until no significant differences in interpretation emerged. 

At first, each author extracted data in sets of six items. During snowballing, the set size was increased to 10.  If the extractor had any doubts about a paper, they were discussed in a meeting. From each set, another researcher verified one data extraction and all exclusions. Any disagreements were resolved by discussion between the extractor and the verifier.

\subsection{Data Analysis}
\label{sec:data-analysis}
We relied on frequency analysis and thematic analysis to analyze our data and answer our research questions. 

\subsubsection{Study Classification}
For study details, we classified the studies into categories that were determined based on the analysis of the pilot set and discussions during the evaluation of the data extraction form. The categories are shown in Table~\ref{tab:dataextractionform}. Studies were assigned to these categories during data extraction.  The frequency of these was then displayed using different data visualization techniques to answer RQ1. 

\subsubsection{Thematic Analysis}
 For the analysis of mental model descriptions, a modified version of the process recommended by Cruzes and Dybå~\cite{cruzes2011recommended} was used.  An initial codebook was created by inductive coding of the initial result set. Coding was used to identify the main concepts and themes in the data. Initial coding was carried out by two researchers independently. The final codebook was then created through discussion. The final codebook was used to code the snowballing result set.

The results related to mental models were also synthesized using thematic analysis, following the practices recommended by Guest et al.~\cite{guest2012}. Text excerpts related to the results were first coded to identify common aspects of mental models that have been studied, for example, background knowledge or use of beacons in comprehension. Results related to each identified aspect were then coded and synthesized to provide a synthesis of the results on those aspects. Coding was performed by one researcher. The analysis was then evaluated and verified by the entire research team. 

\subsection{Overview of the Result Set}
\begin{figure}[H]
  \includegraphics[width=\textwidth]{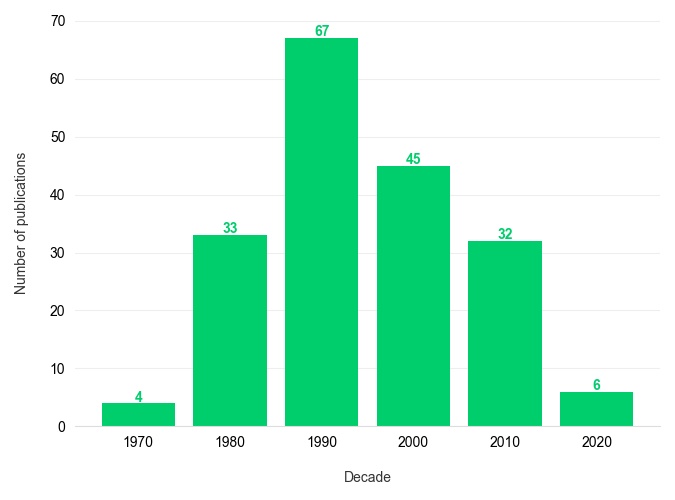}
  \caption{Publications per decade, with the rightmost bar showing publications in the first half of the year 2020. }
  \label{fig:publications_per_decade}
\end{figure}

Our final result set contained 187 results published between 1977 and 2020. Items included in the final result set are listed in~\ref{ap:items}.
The result set contained 88 journal articles, 59 conference papers, 23 workshop or meeting publications, 14 chapters in edited collections or academic books, 2 technical reports, and one pre-print article. 
Of the 187 results, 166 reported a participant study. 
The distribution of the articles over the decades is shown in Figure~\ref{fig:publications_per_decade}. A total of six articles from 2020 met our inclusion criteria. However, we ran the searches in July 2020. This leaves out some of the publications published later in the year.

%% file: 41-results_RQ1.tex
\subsection{Target Systems and Participants}
\begin{figure}[H]
  \centering
  \includesvg[width=\columnwidth]{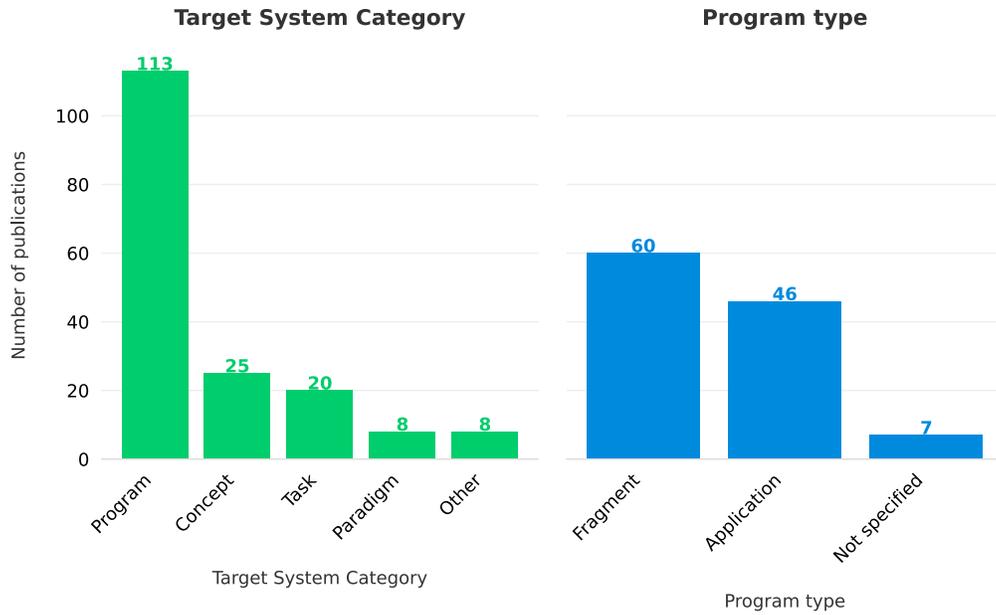}
  \caption{Number of results in each target System Category and Program type  Category \label{fig:target_system_category}.}
\end{figure}

We classified the studies into five target system categories.  The ``Program'' category was further divided into two subcategories. We describe these categories and subcategories in Table~\ref{tab:dataextractionform}. 
Figure \ref{fig:target_system_category} shows the number of articles in each target system category and subcategory. 
Our results show that existing research has been heavily focused on mental models of programs.   
 
\begin{figure}[H]
  \centering
  \includegraphics[width=\textwidth]{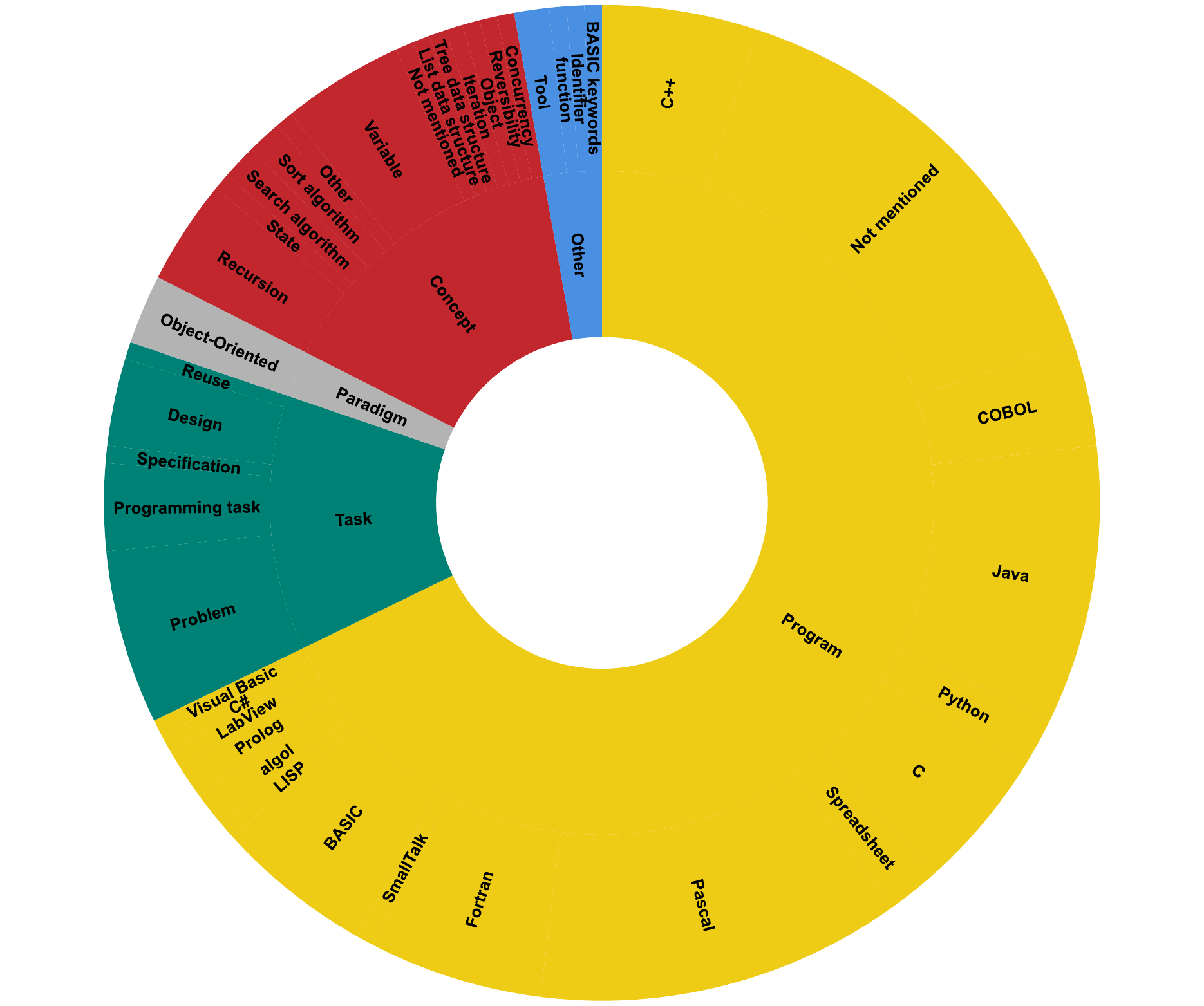}
  \caption{Proportions of different target system types\label{fig:target_systems}.}
\end{figure}

We performed a closer analysis of the types of target systems within each target system category. The results show that a large portion of the studies used older programming languages such as BASIC, Fortran, or Pascal. The popularity of different types of target systems within each target system category is shown in Figure~\ref{fig:target_systems}.

\begin{figure}[H]
  \centering
  \includesvg[width=\columnwidth]{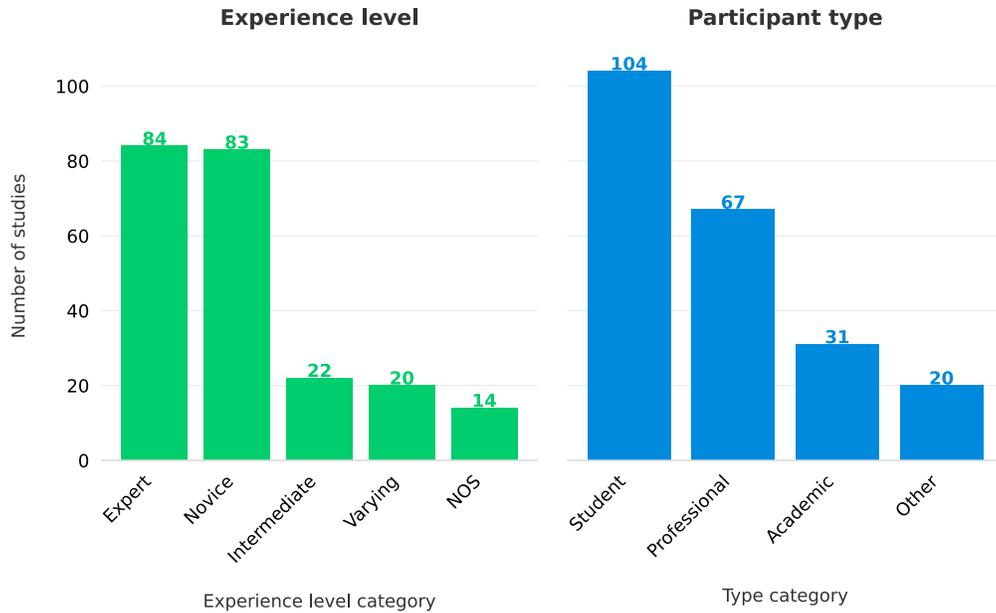}
  \caption{Number of studies in each experience level and participant type category\label{fig:experience_level}.}
\end{figure}

We classified the studies into five categories according to the level of experience of the participants and four categories according to the type of participants. We describe the categories in Table~\ref{tab:dataextractionform}.
Our results show an almost even split between novice and expert participants. However, most of the participants have been students. Figure~\ref{fig:experience_level} shows the number of items in each participant level and participant type category. 

\subsubsection{Evaluating Experience}
63 studies took participant experience into account in their analysis. We evaluated how these studies measured participant experience and the types of experience considered in these studies. 

\begin{figure}[H]
  \centering
  \includegraphics[width=\textwidth]{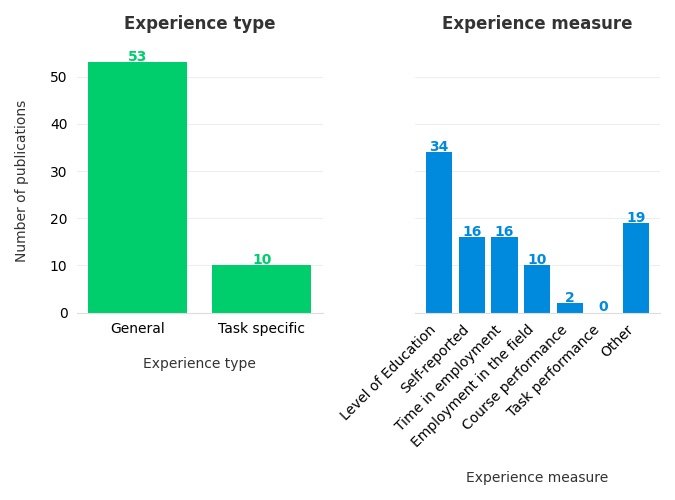}
  \caption{Number of studies considering either experience type (general or task-specific), and number of studies using different experience measures. \label{fig:publications_experience}}
\end{figure}

Most studies considered general computing experience. Only ten of the 63 studies focused on task-specific experience. The studies used different methods to measure experience. The ``other'' category was the second largest, highlighting the use of measures that did not fall under our categorization. In the studies that fit our categorization, the most common measures were indirect measures, such as level of education, time in employment, or self-reporting. Direct measures of skill, such as course or task performance, were rarely used or, in the case of task performance, not used at all. The number of studies that used different types of experience and different experience measures is detailed in
Figure~\ref{fig:publications_experience}.

\subsection{Study Contexts and Study Formats}
We extracted data on study contexts and study formats. We describe the categories used to classify this data in Table~\ref{tab:dataextractionform}. 

\begin{figure}[H]
  \centering
  \includegraphics[width=\textwidth]{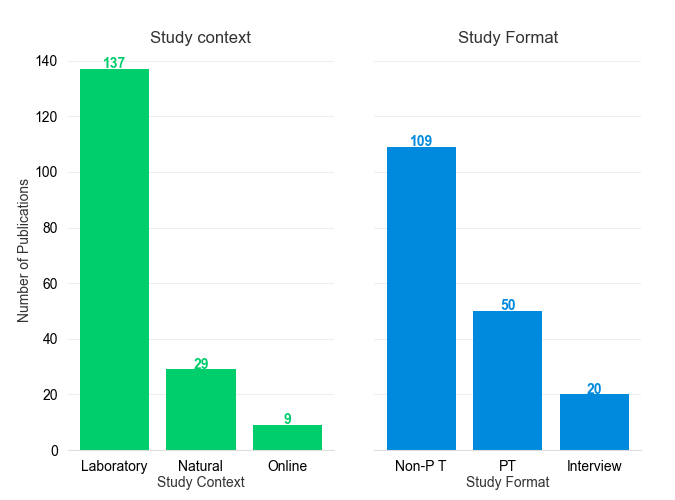}
  \caption{Number of studies in each study format and study context categories. Non-PT = Non-programming task, PT = Programming task. \label{fig:study_format}}
\end{figure}

Most of the studies were conducted in laboratory settings and used non-programming tasks. The number of studies in each category is shown in Figure~\ref{fig:study_format}.

\subsection{Theories}
We identified a set of reoccurring theoretical concepts and analyzed the number of results that mention each concept. We show the theoretical concepts in Table~\ref{tab:dataextractionform}. 

\begin{figure}[H]
  \centering
  \includegraphics[width=\textwidth]{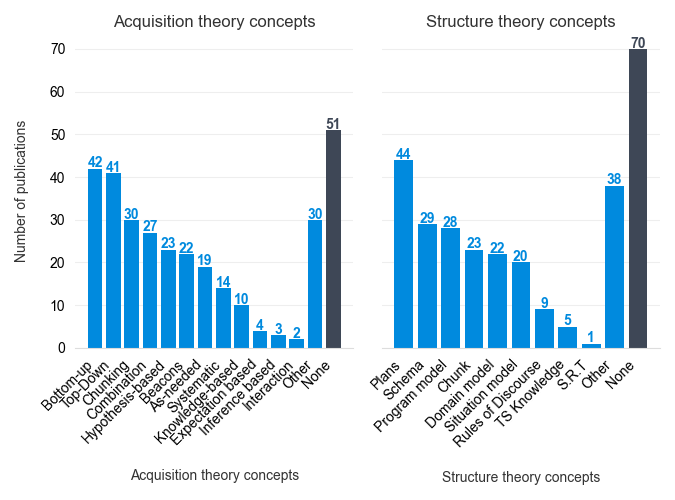}
  \caption{Number of studies mentioning each acquisition and structure theory concept. TS-Knowledge = Text Structure Knowledge.  \label{fig:theory_bar}}
\end{figure}

Several studies did not mention any theoretical concepts. Furthermore, the spread of different theories within this field is wide. Many studies mentioned concepts that we did not see in other studies. These concepts were assigned to the ``other'' category. The number of studies that mention each theoretical concept is detailed in Figure~\ref{fig:theory_bar}.

%% file: 42-results_RQ2.tex
Our analysis of the mental model descriptions revealed three overarching themes in the descriptions. The first theme, overview of mental models, provides an overview of mental models. The second theme, defining characteristics of mental models, describes various characteristics of mental models and the information contained in them. The third theme, mental model acquisition, describes the process of acquiring mental models. The aspects contained within each of the themes are described in Table~\ref{tab:aspects}. 

\renewcommand{\arraystretch}{1.6}
\begin{table}
\begin{tabular}{ p{30mm} p{90mm}}
\textbf{Theme} & \textbf{Description} \\
\toprule
\rowcolor{Gray}
\multicolumn{2}{l}{\textit{Overview of mental models}} \\
\textbf{Definitions} & Nature of MM as cognitive or thought phenomena \newline Areas of memory related to MM \\ 
\textbf{Types of TS} & Concepts, systems, and
situations MM represent \\
\textbf{Uses\newline and Influence} &  Purpose of MM \newline including how MM are used in cognitive processing\newline and their influences on task performance \\
\rowcolor{Gray}
\multicolumn{2}{l}{\emph{Defining characteristics of mental models}} \\
\textbf{Information \newline content} &  Types of information about TS that MM contain. \\
\textbf{General \newline Characteristics} & Abstract characteristics of MM \\
\textbf{Structural \newline characteristics} & Structure of MM \newline Information organization within MM. \\
\rowcolor{Gray}
\multicolumn{2}{l}{\emph{Mental model acquisition}} \\
\textbf{Acquisition} &  Nature of the acquisition process \\
\textbf{Acquisition \newline contexts} &  Contexts in which MM are acquired \\
\textbf{Information sources} &  Information sources used in MM acquisition \\
\bottomrule
\label{tab:aspects}
\end{tabular}
\caption{The aspects within each theme identified in mental model description analysis. MM = mental model, TS = target system}
\end{table}

\subsection{Overview of Mental Models}
\subsubsection{Mental Model Definitions}
Common ways to define mental models were \emph{cognitive structures} or \emph{conceptualizations of thought}. Cognitive structures include concepts such as internal models \cite{wiedenbeck1993characteristics} and internal representations \cite{radermacher2012assigning, gotschi2003mental, abid2019using, sulir2016recording, wiedenbeck1999novice, teasley1994effects, eberts1990mental}. Conceptualizations of thought include definitions such as abstraction of knowledge \cite{armstrong2008building} and characterization of understanding \cite{sorva2008same, sorva2007students, dasgupta2010not, scholtz2010mental, rornes2019students}.

Some studies do not take a position on areas of memory associated with mental models. Some studies describe them as related to short-term memory or working memory \cite{von1994comprehension, abid2019using, detienne1990expert, aziz2014problem, siddiqi1988models}. 
\subsubsection{Types of Target Systems}
 The most common target systems were \emph{systems or programs}. This includes the source code, the underlying machine, and various aspects of a program or a system \cite{sulir2016recording, corritore1999mental, navarro2001visual, wiedenbeck1999novice, ramalingam1997empirical, teasley1994effects, corritore1991novices, pennington1987stimulus, yates2020characterizing, burkhardt1997mental, wiedenbeck1993characteristics, boehm1996techniques,detienne1990empirically, karahasanovic2007comprehension, von1994dynamic, romero2001focal, sillito2008asking, vainio2007factors}.
 
Target systems also included \emph{programming concepts} \cite{benvenuti2015representation, sajaniemi2005investigation, radermacher2012assigning, gotschi2003mental, scholtz2010mental, pieterse2018triviality, sheetz2002identifying, kahney1983novice, rornes2019students}, and \emph{programming situations}. This includes programming tasks and solutions \cite{siddiqi1988models, davies1990plans, gilmore1988programming, balijepally2012effect, shaft2006role, koubek1988theory, fritz2014developers, shaft2006role}, and programs being implemented or designed \cite{molzberger1986analyzing, dawson2013cognitive, burkhardt1995empirical, koubek1988theory}. 
\subsubsection{Uses and Influence}
Some studies describe mental models as mental representations that allow one to \emph{derive information} about the target system through mental simulation \cite{petre1997glimpse, hoc1977role, adelson1985comparing} and reasoning \cite{radermacher2012assigning, dawson2013cognitive, wiedenbeck2004factors}. Some studies describe them as representations that, in general, help \emph{interpret and explain} the target system \cite{balijepally2015task, balijepally2012effect, chao2018dynamic, wiedenbeck2004factors,eberts1990mental}.

Mental models have been described as aiding in performing tasks. These tasks include decision making \cite{eberts1990mental, balijepally2015task}, making predictions \cite{scholtz2010mental, dawson2013cognitive, hoc1977role, balijepally2015task, chao2018dynamic}, and various programming tasks \cite{davies1990plans, petre1997glimpse, aziz2014problem} such as debugging \cite{eberts1990mental, wiedenbeck1993characteristics} and program modification \cite{wiedenbeck1993characteristics, karahasanovic2007comprehension, corritore2000direction}.

Some articles mention that mental models influence task performance. They mention that mental models help perform tasks, for example, by reducing the cognitive load during program comprehension\cite{bisant1993cognitive}.

\subsection{Defining Characteristics}
\subsubsection{Information Content}
Some articles described mental models as containing \emph{knowledge and information} related to the target system \cite{wiedenbeck1999novice, teasley1994effects, wiedenbeck1993characteristics, detienne1990empirically, letovsky1987cognitive, wiedenbeck2004factors, boehm1996techniques, letovsky1987cognitive, navarro1999mental, detienne1990empirically, gilmore1988programming}. Others include \emph{strategies} to comprehend the system and \emph{background knowledge} related to the system and the situational context as parts of the mental model~\cite{wiedenbeck2004factors}.

Some articles specify that the knowledge in mental models represents either the structure or function of the target system or both. \emph{Structure} often refers to parts of the target system and how they are connected \cite{corritore1991novices, fritz2014developers, altmann1999episodic, gilmore1988programming, balijepally2012effect, balijepally2015task, siddiqi1988models, wiedenbeck2004factors}. When discussing abstract conceptual target systems such as programming concepts, the structure can also refer to concepts related to the target system and the relationships between them \cite{armstrong2007understanding}. \emph{Function} refers to knowledge of how and why the target system works \cite{corritore1999mental, wiedenbeck1999novice, yates2020characterizing, detienne1990empirically, karahasanovic2007comprehension, navarro1999mental, pieterse2018triviality, rornes2019students, wiedenbeck2004factors}.
\subsubsection{General Characteristics}
Some studies say that mental models are of \emph{varying quality}. This refers to descriptions of mental models as representing the reality of the target system with varying accuracy. Mental models are also described as varying in viability \cite{gotschi2003mental, sorva2007students}, completeness \cite{romero2001focal, chao2018dynamic}, and strength \cite{detienne1990empirically, gilmore1988programming}. Mental models have also been described as something that can be naive or based on incorrect assumptions.

Mental models are described as \emph{dynamic}. This includes notions of changing and updating mental models, as well as the notion of mental models as temporary \cite{abid2019using, balijepally2015task}.

 Mental models are also said to be affected by the characteristics of the comprehension situation \cite{romero2001focal, balijepally2015task} or activated by the specific situation. They have been described as subjective \cite{hoc1977role} and different between novices and experts \cite{fix1993mental, corritore1991novices, romero2001focal, kahney1983novice}. However, some studies equally describe mental models as similar between programmers.
 
Other attributes of mental models present in the data include runnable \cite{vainio2007factors} , hierarchical \cite{romero2001focal, fix1993mental, boehm1996techniques, shargabi2015program, jeffries1982comparison}, and as describing something abstract.
\subsubsection{Structural Characteristics}
The structure of mental models has been described in two ways. A recurring theme is mental models as a layered structure that provides alternative views of the target system at different levels of abstraction \cite{poruban2015program, khazaei2002there, von1994comprehension, corritore1999mental, navarro2001visual, ramalingam1997empirical, bratthall2001program, burkhardt1997mental, letovsky1987cognitive, von1994dynamic, von1996identification, shargabi2015program, anneliese1998program, jeffries1982comparison, von1993code, von1996identification, von1996roleunderstanding}. Another recurring theme is mental models as organized collections of knowledge, describing mental models as containing or consisting of collections, groups, or clusters of related knowledge organized according to some criteria or criterion \cite{romero2001focal, pennington1987stimulus, wiedenbeck1993characteristics}.

Mental models are also described in terms of their relationships with the target system, and some descriptions highlight the mapping between the target system and the mental model or the mappings between the different layers of abstraction present in the model \cite{fix1993mental, letovsky1987cognitive, vainio2007factors}.

\subsection{Mental Model Acquisition}
\subsubsection{Acquisition Contexts}
Acquisition contexts described in the literature include different learning and sense-making situations. The literature describes educational settings~\cite{radermacher2012assigning, stromback2019student,sorva2007students}, task contexts such as program design~\cite{sulir2016recording, kim1995internal}, and sense-making situations such as learning to use a new system~\cite{bayman1983diagnosis}.
\subsubsection{Acquisition Process}
A common feature of mental model acquisition is the notion of \textit{activation of background knowledge}. This refers to the features of the target system activating the programmer's background knowledge of related concepts or programming patterns~\cite{yates2020characterizing, kulkarni2017supporting, davies1995objects, detienne1990program, soloway1984empirical, wiedenbeck1991initial}.

Two specific acquisition situations have been described in the literature, \textit{program comprehension}~\cite{fix1993mental,bidlake2020systematic,navarro2001visual,ramalingam1997empirical, detienne2002understandingeffects} and \textit{interaction with the target system}~\cite{eberts1990mental, norman1983some}.
\subsubsection{Information Sources}
Mental models were commonly described as being informed by the combination of information within the environment and the skills and experiences of the programmer \cite{benvenuti2015representation,shaft2006role}. 
They were commonly described as being informed by the source code of a program \cite{fix1993mental, wiedenbeck1993characteristics, buckley2004characterizing} or, more specifically, the structure of the program or the recurring patterns found within the source code \cite{fix1993mental, kulkarni2017supporting}.
 
 Supporting material such as documentation and task specifications was also referenced in the literature \cite{shaft1995relevance, eberts1990mental, wiedenbeck1993characteristics}.
 
The programmer's skills and experiences were also mentioned. Studies stated that mental models are informed by the programmer's background knowledge, skill level, and experiences \cite{shaft2006role,sheetz2002identifying, yates2020characterizing, kim1995internal, robertson1990knowledge, soloway1988knowledge, xu2005cognitive, von1995program}.

%% file: 43-results_RQ3-4.tex
\begin{figure}[H]
  \centering
  \includegraphics[width=\textwidth]{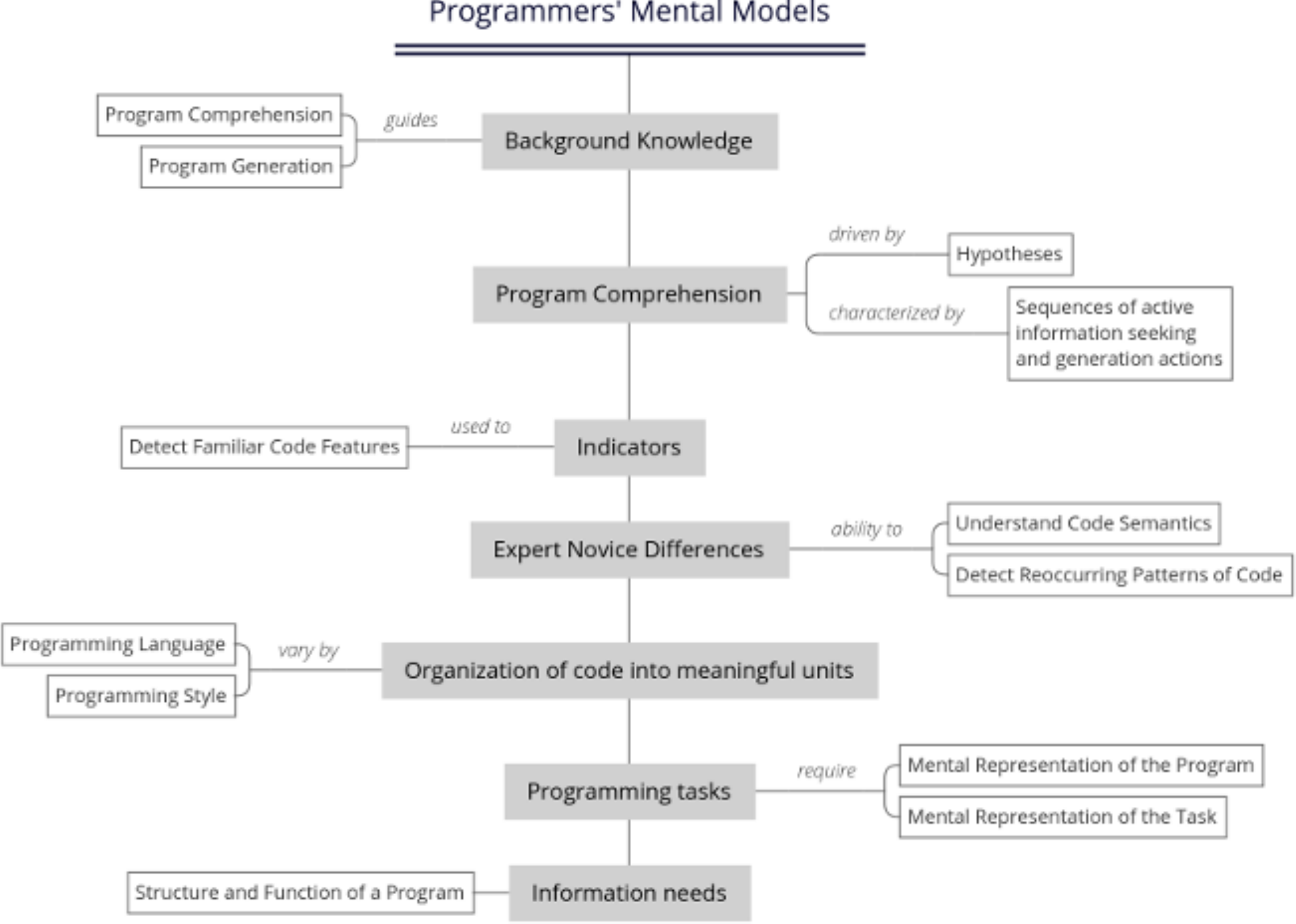}
  \caption{Overview of the main characteristics and attributes of mental models and their acquisition\label{fig:mm_overview}}
\end{figure}

With RQ3 and RQ4, we synthesized the existing knowledge on programmers' mental models by uncovering recurring themes and concepts in the results related to mental models. Figure~\ref{fig:mm_overview} displays the aspects and characteristics discussed in this section.

\subsection{Background Knowledge Guides Program Comprehension and Program Generation}
\subsubsection{Types of Background Knowledge}
Programmers have \emph{background knowledge} of programs from different domains, programming concepts, familiar programming sequences, syntax of programming languages, and common syntax and formatting rules used in programming \cite{poruban2015program, fix1993mental, pennington1987stimulus, green1995programming, detienne1990empirically, letovsky1986delocalized, letovsky1987cognitive, von1993code, von1994comprehension, von1995industrial, von1996identification, o2004expectation, robertson1990knowledge, shaft1995relevance, wiedenbeck1991initial, yu1988plan, romero2001focal, romero2004structural, davies1990plans, gilmore1988programming, soloway1984empirical, soloway1988knowledge, adelson1981problem, detienne1990expert, von1994dynamic, von1995program, schomann1994knowledge, davies1995objects, ikutani2021expert, stromback2019student, masmoudi2001solving, bonar1985preprogramming, davies1990plans, rist1991knowledge, scholtz1996adaptation}. We identified the following types of background knowledge that have been studied:

\begin{bknowledge}
\item Knowledge of \emph{programming concepts} refers to knowledge of abstract programming concepts such as recursion \cite{sajaniemi2005investigation, gotschi2003mental, scholtz2010mental,pieterse2018triviality}.
\item Knowledge of \emph{programming patterns} or plans refers to knowledge of reoccurring code sequences that implement some functionality \cite{detienne1990expert, von1993code,von1994comprehension, von1995program, von1996roleunderstanding, letovsky1987cognitive, romero2004structural, green1995programming, adelson1981problem, davies1990plans, rist1991knowledge, scholtz1996adaptation, ehrlich1984empirical}.
\item Knowledge of the programming language \emph{syntax} refers to knowledge of how these abstract concepts and patterns are implemented in different programming languages \cite{romero2004structural, romero2001focal, robertson1990common, fix1993mental, detienne1990expert, gilmore1988programming, davies1990plans, shaft1995relevance, soloway1984empirical, soloway1988knowledge, wiedenbeck1991initial, letovsky1987cognitive, yu1988plan, pennington1987stimulus, bayman1983diagnosis}.
\item  \emph{Domain knowledge} represents knowledge of common architectures of programs in different domains \cite{jeffries1981processes} and programming concepts related to these domains \cite{detienne1990empirically, o2004expectation, armstrong2007understanding}.
\item Knowledge of abstract \emph{high-level programming concepts} represents programmers' perception of  abstract concepts such as encapsulation, coupling \cite{benvenuti2015representation, bavota2013empirical, voigt2009intuitiveness}, or object-oriented programming \cite{hubwieser2011investigating, sheetz2002identifying, cooke1988effects}.
\end{bknowledge}

\subsubsection{Attributes of Background Knowledge}
Programmers' background knowledge is described as having certain reoccurring attributes. An attribute that has been highlighted especially when studying novices is \emph{consistency}. Consistency refers to the ability to generalize conceptual knowledge across situations and recognize the same abstract concept in multiple concrete code representations \cite{caspersen2007mental, bornat2008mental, radermacher2012assigning}. We can also think of conceptual knowledge in terms of \emph{viability}, or how accurately and completely mental models represent the reality of the target systems \cite{ma2007investigating, gotschi2003mental, scholtz2010mental, pieterse2018triviality, bayman1983diagnosis, sorva2007students, sanders2012first, rornes2019students}.

\subsubsection{Acquisition of Background Knowledge}
The acquisition of background knowledge has been discussed in the literature. Understanding of programming concepts appears to be acquired through \emph{education}. In computer science education, students are taught concepts such as variable assignment \cite{ma2007investigating, pieterse2018triviality}, recursion \cite{gotschi2003mental, kessler1986learning, sanders2012first}, concurrency \cite{stromback2019student}, and state \cite{herman2017affordances}. Students also learn of higher-level abstract concepts such as object-oriented programming \cite{hubwieser2011investigating, sanders2008student}.

Studies suggest that knowledge of reoccurring programming sequences develops through \emph{practice}, and is acquired as the programmer develops solutions to different programming problems \cite{rist1991knowledge, scholtz1996adaptation, ehrlich1984empirical}.

In some situations, it seems that conceptual and plan knowledge are transferred from one domain to another. Conceptual knowledge from, for example, the domain of mathematics can be applied to understanding code, and knowledge from one programming domain can be transferred to another \cite{gotschi2003mental, pieterse2018triviality, armstrong2007understanding}. The same holds when transferring between programming languages, where programming plans are adapted from one language to another \cite{scholtz1996adaptation}.
\subsubsection{Use of Background Knowledge in Program Comprehension and Program Generation}
Studies have investigated how programmers use their background knowledge to complete programming-related tasks~\cite{letovsky1986delocalized,soloway1988knowledge,detienne1990empirically}. These studies show that background knowledge is an intrinsic part of task performance. A key observation is that programmers use their background knowledge to make sense of new systems and situations and to construct new code solutions \cite{soloway1988knowledge, letovsky1986delocalized,poruban2015program, pennington1987stimulus, robertson1990knowledge, yu1988plan, fix1993mental, green1995programming, schomann1994knowledge, soloway1984empirical, shaft1995relevance, detienne1990empirically, detienne1990expert, ikutani2021expert, davies1995objects, kahney1983novice}.

Background knowledge allows \emph{mapping} between code and semantic knowledge of the program function and, therefore, allows detecting and interpreting meaningful code structures \cite{soloway1988knowledge, schomann1994knowledge, soloway1984empirical, shaft1995relevance, detienne1990empirically, detienne1990expert}. It helps to establish expectations of the programs and therefore drives program comprehension by compelling programmers to seek information that matches those expectations \cite{soloway1984empirical, von1994comprehension, detienne1990empirically, von1994dynamic, o2004expectation}.
Knowledge of programming concepts and patterns also allows reasoning according to and about code structures during program comprehension \cite{letovsky1987cognitive, detienne1990empirically}.
Background knowledge can also help to decide which aspects or elements of the program are relevant to understand to complete the task at hand \cite{altmann1999episodic}.

Knowledge of programming concepts and patterns is also used in program generation and other related tasks such as debugging \cite{kahney1983novice}. Conceptual knowledge of concepts such as recursion helps programmers implement solutions using those concepts \cite{kahney1983novice, bonar1985preprogramming, rist1991knowledge, whalley2014qualitative}. Reoccurring programming patterns are also used to generate programs. These patterns can be retrieved from memory as required by the task and then used to guide the formation of a solution \cite{hoc1977role, adelson1985comparing, bonar1985preprogramming, rist1991knowledge, whalley2014qualitative, ehrlich1984empirical}.
Domain knowledge has also been shown to help in programming tasks and especially in program design \cite{adelson1985comparing, jeffries1982comparison}. Knowledge of common concepts, patterns, and architectures within a domain appears to aid in the design of new programs for that domain \cite{adelson1985comparing, aziz2014problem, jeffries1982comparison}. 

\subsection{Program Comprehension Is Driven by Hypotheses}
Programmers have been observed generating \emph{hypotheses} about the features, structure, and function of a program in many studies \cite{poruban2015program, von1994comprehension, von1995program, bratthall2001program, o2004expectation, koenemann1991expert, xu2005cognitive, anneliese1998program, von1993code, von1996identification, von1997program, votipka2020observational, sutcliffe1992analysing}.
Different levels of hypotheses have been observed.
High-level hypotheses about the structure and functional elements of a program are used to guide further comprehension efforts \cite{bratthall2001program, von1997program, votipka2020observational}. High-level hypotheses about the function of the program have also been observed \cite{anneliese1998program, votipka2020observational}.
Hypotheses about which code areas are relevant for the task at hand \cite{koenemann1991expert, xu2005cognitive, votipka2020observational} and where these are located \cite{anneliese1998program, jeffries1982comparison} guide the programmers' investigation efforts towards the relevant areas.
Lower-level hypotheses about individual program elements or functionalities drive the search for confirmatory features \cite{arunachalam1996cognitive}.

Hypotheses can be in the form of \emph{expectations}.  Programmers build expectations about code based on their prior knowledge and then seek code features to confirm their hypotheses \cite{o2004expectation, latoza2007program}.
On the other hand, hypotheses can be \emph{inference based}, or assumptions about some feature encountered in the code formed from its indicators \cite{o2004expectation, latoza2007program}. These are hypotheses about what the code artifact indicated by the indicators is or does and are answered by examining the artifact to confirm or reject the hypothesis \cite{o2004expectation, latoza2007program}.

Familiarity with the domain results in more high-level expectations, i.e., the presence of domain schemata or plans leads to more top-down processing \cite{o2004expectation, shaft1995relevance, von1997program, vessey1996computer}.

\subsection{Indicators Are Used to Detect Familiar Code Features}
\emph{Indicators} refer to code or environmental features that provide clues to some functionality or structure present in the code. In the literature, these have often been referred to as beacons and cues, although the vocabulary varies between studies.
Indicators indicate the presence of familiar structures and can be used to detect the presence of familiar structures or semantic elements from code \cite{davies1995objects, schomann1994knowledge, wiedenbeck1991initial, o2004expectation, teasley1994effects, siegmund2017measuring, masmoudi2001solving}.
\subsubsection{Types of Indicators}
The main types of indicators present in the literature are \emph{semantic hints} and \emph{focal lines}.
Semantic hints are indicators such as meaningful variable, function, and file names, comments, and other elements that provide semantic hints to program functionality \cite{crosby2002roles, teasley1994effects, hofmeister2017shorter, bratthall2001program, gellenbeck1991investigation, avidan2017effects, koenemann1991expert, levy2019understanding}. They allow for building expectations of code and interpreting it \cite{crosby2002roles, takang1996effects, koenemann1991expert}.
The focal lines represent the lines of code most indicative of some familiar programming sequence and can be used to recognize those programming sequences from code \cite{soloway1984empirical, wiedenbeck1991initial, crosby2002roles, romero2004structural}.
\emph{Code conventions} can also be understood as indicators. With code conventions, we refer to customary ways of commenting, naming, and structuring code that help to build expectations about the code and recognize program structures \cite{levy2019understanding}.
\subsubsection{Use of Indicators in Program Comprehension}
In program comprehension indicators have been shown to benefit the comprehension process in two ways.
Indicators that suggest that a familiar structure is present in the code can provide the basis for establishing expectations about the code, guiding further program comprehension efforts to confirm those expectations \cite{schomann1994knowledge, letovsky1987cognitive, o2004expectation, gellenbeck1991investigation, poruban2015program}.
Indicators can also be used to confirm hypotheses by confirming that indicators indicating that the hypothesis is correct are present in the code \cite{schomann1994knowledge, shaft1995relevance, gellenbeck1991investigation}.
Large-scale programs also span multiple files, and beacons can be used to denote the locations of code elements in the codebase \cite{bratthall2001program}.

The ability to utilize and detect indicators in code may depend on the programmer's level of experience. Especially focal lines seem to be better detected by more experienced programmers \cite{davies1994knowledge}.
At least for novices, the usefulness of indicators may vary according to how well the indicators match the programmer's understanding of the concept or functionality the indicators indicate~\cite{chao2018dynamic}.

Inferences made from indicators are not always accurate. Assumptions made about the code based on indicators \cite{poruban2015program, latoza2007program} are not necessarily investigated or verified.  This can be seen in situations where false or misleading indicators lead to a false understanding of the code \cite{wiedenbeck1991initial, avidan2017effects}.

\subsection{Program Comprehension is Characterized by Sequences of Active Information Seeking and Generation Actions}
Studies have followed programmers' \emph{comprehension processes} in naturalistic program comprehension settings.
These studies have shown program comprehension as an active process of \emph{seeking information} of the program being comprehended driven by the programmer's \emph{goals} and \emph{information needs}. The program comprehension process is made up of sequences of actions that are intended to gather information to understand some feature(s) or area(s) of a program according to goals and information needs \cite{von1994comprehension, von1994dynamic, von1996identification, burgos2007through, anneliese1998program, ko2003individual, latoza2007program, von1997program, votipka2020observational, letovsky1987cognitive, sharpe1997unifying, von1996roleunderstanding}.
Whether this process is similar to reading text has been debated, but some studies suggest that it uses different brain networks \cite{liu2020computer}.
\subsubsection{Information Needs and Sources}
Although all information-seeking goals are geared toward gathering some aspect of information, the types of information vary. Observed in the literature are understanding the purpose or function of an encountered feature \cite{anneliese1998program, von1996identification}, understanding how a functionality is implemented \cite{sillito2008asking, anneliese1998program},
understanding what a code element is \cite{anneliese1998program}, understanding how the code behaves \cite{letovsky1987cognitive, sillito2008asking}, understanding what elements are relevant to the task at hand and where they are located \cite{sillito2008asking, latoza2007program}, and understanding how those elements are connected to other elements of the program \cite{sillito2008asking}.

The program comprehension process uses information from multiple \emph{information sources} both within the code itself and from other sources such as the Internet, documentation, and other programmers \cite{burgos2007through, karahasanovic2007comprehension, koenemann1991expert, levy2019understanding}.

\subsubsection{Program Comprehension Actions}
The program comprehension process utilizes both hypothesis-based \emph{top-down} processes and inference-based \emph{bottom-up} processes \cite{sharpe1997unifying, von1994comprehension, von1994dynamic, anneliese1998program} switching between methods based on the programmer's current knowledge and information needs for the current goal.

Different \emph{information gathering actions} have been documented. Some of these actions include interacting with the code to extract and comprehend the information present in it. These actions include:
\begin{inparaenum}[i)]
\item Reading code line by line \cite{anneliese1998program, von1997program};
\item mentally simulating code execution at various levels of abstraction to understand code execution and interactions and relationships between code elements \cite{burgos2007through, detienne1990empirically, anneliese1998program, fleming2008refining, jeffries1982comparison, detienne1990expert, latoza2007program};
\item reasoning about code based on prior knowledge \cite{detienne1990program, latoza2007program} to form hypotheses and explain or interpret code that has been encountered;
\item running the code to examine it at run-time \cite{karahasanovic2007comprehension, levy2019understanding, votipka2020observational};
\item use of a debugger to examine code execution \cite{burgos2007through, sillito2008asking}.
\end{inparaenum}

Some of the actions include searching for code elements to comprehend. These actions include searching for specific code elements to examine \cite{burgos2007through, sillito2008asking, votipka2020observational} and scanning the code to find specific code elements or indicators \cite{sillito2008asking, jeffries1982comparison, votipka2020observational, bratthall2001program}.

\subsubsection{Factors Affecting Program Comprehension}
The comprehension process varies in \emph{scope}, or how much of the codebase is comprehended \cite{von1994comprehension, corritore2001exploratory, karahasanovic2007comprehension, koenemann1991expert, fleming2008refining, szabo2015novice, fritz2014developers}, and \emph{direction}, or if the comprehension process proceeds from a higher-level hypothesis to seeking information or from comprehending smaller code structures to building a higher-level hypothesis \cite{levy2019understanding, torchiano2004empirical, szabo2015novice}.
Some factors that affect the process include the programmer's goals for the comprehension situation \cite{von1994comprehension, koenemann1991expert}, the type of code being comprehended, the programmer's overall level of programming experience \cite{eberts1990mental}, and their experience with different types of code \cite{boehm1996techniques, corritore2001exploratory, corritore2000direction}, as well as the task at hand \cite{parkin2004exploratory, szabo2015novice}.
Other individual characteristics of the programmer and the situation, such as the use of a screen reader \cite{armaly2017comparison} or contextual features such as code ownership, can also affect the comprehension process \cite{dasgupta2010not}. 

\subsection{Structure and Function of a Program Are Prevalent Information Needs}
At a higher level, important aspects of a program to understand seem to be its structure and function, and at a lower level, the implementation of the said function \cite{levy2019understanding, yates2020characterizing}. Programmers have been shown to seek both implementation-level information and abstract functional and structural information \cite{bratthall2001program, von1994comprehension, levy2019understanding, von1993code, von1997program, von1996identification}, however, the prevalence of either type of information varies.

With understanding structure and function, we refer to understanding the program architecture, or the composition and interaction of functional elements that comprise the program \cite{yates2020characterizing, sillito2008asking, fleming2008refining}, including the role and function of individual functional elements \cite{levy2019understanding}, the relationships between the various functional elements and their interactions at run-time \cite{yates2020characterizing, sillito2008asking, levy2019understanding, fleming2008refining}.
With understanding implementation, we refer to a more comprehensive understanding of how any individual feature or functionality is implemented \cite{yates2020characterizing} and behaves during execution \cite{fleming2008refining, levy2019understanding}.

Although structure, function, and implementation are the most prevalent information categories present in the literature, other information categories were also identified. Understanding the rationale behind why the program is implemented as is is a recurring theme in the literature \cite{yates2020characterizing, levy2019understanding, burgos2007through}.
Furthermore, the temporal understanding of a program or the understanding of past and ongoing development activities and changes in a program appears to be relevant in some situations \cite{yates2020characterizing, levy2019understanding}.
In studies on the comprehension of systems programmers work with for longer periods of time, another aspect of information is the notion of the use of functional elements, or understanding when and how certain elements should be used in modification and enhancement tasks \cite{levy2019understanding}.

The abstraction level at which the program is understood varies.
The task to do with the code affects the level of understanding required. Studies have shown that the level of abstraction at which code is understood depends on the task requirements \cite{burkhardt2002object, levy2019understanding}.
Knowledge at the implementation level appears to be acquired only as needed and not always required \cite{yates2020characterizing, levy2019understanding}.
The level of understanding required for the task may also depend on the difficulty of the task \cite{boehm1992role}.

\subsection{Novice and Expert Programmers Differ in Their Ability to Understand Code Semantics and Detect Reoccurring Patterns of Code}
Novice and expert programmers differ in their ability to understand programs and generate code.
\subsubsection{Abstraction}
One critical difference between experience levels may stem from differences in the ability to understand programs, both existing programs and programs being designed, at a more abstract level.
Although the results vary between studies \cite{scholtz1996adaptation}, most of the research points to the fact that novice programmers struggle to understand the abstract semantic features of programs more than experienced programmers \cite{burkhardt2002object, corritore1991novices, burkhardt1997mental,shargabi2015program, jeffries1982comparison, latoza2007program, adelson1984novices}. Novices seem to struggle to build a higher-level abstract representation of the structure and function of programs \cite{burkhardt1997mental, wiedenbeck1986organization, ye1996expert, jeffries1982comparison, corritore1991novices, vainio2007factors}.
We observe the same phenomenon in program design, where novice programmers struggle to construct abstract mental representations of the programs they are designing \cite{adelson1985role}. Some studies show that novice mental models of code also vary in accuracy, that is, they vary in how accurately they represent the different features of code \cite{ramalingam2004self}.
\subsubsection{Differences in Background Knowledge}
These differences in the ability to understand program semantics may stem from differences in background knowledge. Novice programmers show less knowledge of code patterns, which could hinder their ability to match semantic information with code or problem representations \cite{latoza2007program, teasley1994effects, corritore1991novices, karahasanovic2007comprehension, adelson1985comparing, rist1991knowledge}. Novices' knowledge appears to be more syntactically based \cite{adelson1981problem, mckeithen1981knowledge, wiedenbeck1986organization}. Experts, on the other hand, seem to think of concepts they detect in code in more abstract and semantic terms \cite{sajaniemi2005investigation, eberts1990mental}.

Furthermore, novices have misconceptions about the semantics of familiar programming concepts \cite{jeffries1981processes, jeffries1982comparison} and can have misconceptions about the functionality of programming statements \cite{bayman1983diagnosis}. These misconceptions may affect their ability to understand a program \cite{sajaniemi2008study}.
\subsubsection{Ability to Recognize Reoccuring Patterns}
Experts' code comprehension performance may be partly due to their ability to detect familiar patterns, as their performance also deteriorates when the code is not in expected order, making this recognition of familiar patterns more difficult \cite{detienne1990expert, bateson1987cognitive}. In some studies, this is also supported by differences between novices and experts in the use of top-down versus bottom-up methods, showing that novices rely more frequently on bottom-up reading \cite{abid2019using}, although there are contrasting results showing that novices use top-down methods \cite{ye1996expert}.
Experts have been shown to detect familiar programming patterns in code better than novices \cite{fix1993mental, jeffries1982comparison, latoza2007program, barfield1986expert, barfield1997skilled, mckeithen1981knowledge, schomann1994knowledge, boehm1996techniques, ikutani2021expert, teasley1994effects}, and demonstrate a better ability to detect indicators in code \cite{teasley1994effects, latoza2007program, crosby2002roles}. Experienced programmers also have a better ability to utilize reoccurring programming patterns in program generation \cite{davies1990plans, davies1991role, jeffries1982comparison}.

However, the differences between experts and novices in program comprehension cannot be fully explained by differences in pattern recognition. Novices also struggle with other comprehension activities, such as tracing code execution \cite{vainio2007factors}, reading code in sensible order \cite{jeffries1982comparison}, and finding relevant code structures \cite{jeffries1982comparison}. Reasoning skills appear to also develop with more experience \cite{izu2017ability}.

\subsection{Different Programming Styles and Programming Languages Support Different Organization of Code into Meaningful Units}
Code comprehension is based on the detection and comprehension of meaningful code units \cite{barfield1986expert, barfield1997skilled, mckeithen1981knowledge, wiedenbeck1986organization}.
Programmers detect code segments that represent some meaningful entity, and the detected entities seem to overlap between programmers, suggesting that some more universal basis is used to group these elements \cite{sulir2016recording, davies1995objects}.

However, the code representations of the abstract elements vary between programming languages, and the code features used to detect these elements also vary \cite{khazaei2002there}. Due to this, the reliance on different aspects of code, such as data structures or functions, in the detection and extraction of code elements varies between programming languages \cite{romero2004structural, navarro2001visual, navarro1999mental, green1995programming}.

Different programming styles also highlight different aspects of the code. They organize the code into meaningful groups, elements, and functions in different ways, therefore, supporting the understanding of different aspects of the code \cite{khazaei2002there, wiedenbeck1999novice, ramalingam1997empirical, navarro1999mental}.
For example, object-oriented programming supports organizing programs into meaningful classes and objects. This supports a more structural understanding of the program \cite{khazaei2002there, corritore1999mental, wiedenbeck1999novice, ramalingam1997empirical}. Other programming styles, such as event-driven or procedural,  support a more detailed understanding of individual code statements and control flow \cite{khazaei2002there, corritore1999mental, wiedenbeck1999novice, ramalingam1997empirical}. However, some studies suggest that the effect of the coding style on the resulting cognitive representations may depend on the level of programming experience of the programmer \cite{boehm1992role}.

\subsection{Programming Tasks Require Mental Representations of the Program and the Task}
One factor in performing programming tasks appears to be the ability to construct mental representations of the coding solution to be implemented \cite{balijepally2012effect, hoc1977role, adelson1985comparing, kim1995internal}.  In the cases where the programming task is done within an existing codebase, this includes the ability to create a mental representation of the program to be modified. In many studies, this also includes a mental representation of the problem or an understanding of the problem for which the solution is implemented \cite{hoc1977role}.
\subsubsection{Mental Models of Programs Being Developed}
Programmers have been observed creating multisensory, dynamic mental representations of programming solutions. The form of these representations seems to be highly individual \cite{petre1999mental, petre1997glimpse, molzberger1986analyzing}. They allow for internal representation of the solution at multiple levels of abstraction \cite{petre1999mental, petre1997glimpse, molzberger1986analyzing}.

These mental models represent the programmer's understanding of the solution or the program to be implemented \cite{adelson1985comparing, kim1995internal} and seem to allow the programmer to mentally simulate the program to allow them to understand how it would function \cite{adelson1985comparing, kim1995internal}.
These models can contain abstractions of key aspects of the new program, such as its main objects \cite{dawson2013cognitive, chatel1994expertise} or functions \cite{chatel1994expertise}. They can also contain representations of the causal relationships between key elements of the program \cite{adelson1985comparing, chatel1994expertise} and the requirements of the system \cite{adelson1985comparing}. In some situations, programmers can also develop mental models of code elements that they would like to reuse to create the new solution code \cite{burkhardt1995empirical}.

\subsubsection{Acquisition of Mental Models of Programs Being Developed}
When designing a new program, mental models can be created based on information on the requirements and technologies of the new program collected from multiple sources, including customers and team members, and the review of similar systems \cite{dawson2013cognitive, aziz2014problem}.
The creation of these mental models of solutions seems to start from an abstract representation of the key aspects of the solution \cite{dawson2013cognitive}. This is then systematically expanded to a more concrete level \cite{adelson1985comparing, kant1984problem, rist1989schema, rist1991knowledge, siddiqi1988models}. Some studies hypothesize that the initial kernel idea comes from background knowledge of similar programs \cite{rist1989schema, rist1991knowledge}, which is then systematically expanded to suit the current problem situation.
\subsubsection{Task Context Models}
In programming situations where the programmer works with an existing codebase, they create mental representations not only of the current programming task but also of the program on which they are working \cite{shaft2006role, koubek1989implementation, sutcliffe1992analysing}. In program design situations, we also see representations of the problem domain, but here in the form of understanding the abstract problem domain where the program will operate \cite{aziz2014problem, sutcliffe1992analysing}. The methods for comprehending programs in different task situations do not seem to differ from general program comprehension methods, which have been discussed in previous sections.
\subsubsection{Effects of Expertise}
In maintenance tasks, novice programmers have shown to have difficulty understanding the program they work on, and this may hinder their performance \cite{boehm1992role}. Similar differences can be detected between experts and so-called super-experts \cite{koubek1989implementation}.

As programming experience increases, so does the ability to represent tasks, programs, and problems as larger abstract units \cite{hoc1977role}. Beginner programmers have been shown to meticulously translate problem statements into code step by step, whereas more experienced programmers show the ability to represent elements of both problems and solutions as larger abstract units \cite{hoc1977role, adelson1985comparing, rist1991knowledge, koubek1988theory}.
When comparing programmers of different experience levels in modification tasks, one of the attributes of an expert programmer remains their ability to create an abstract and complete representation of the program to be modified \cite{eberts1990mental, koubek1991cognitive} and the task at hand \cite{koubek1988theory}.

%% file: 50-discussion.tex
In this section, we discuss the answers to our research questions. Until now, research has focused on examining mental models of programs in laboratory settings. Due to the fragmentation of the field, these results have been scattered and difficult to connect to a coherent whole. However, through our analysis, we were able to show that there is a kernel of shared understanding about programmers' mental models. We were also able to find some aspects of mental models that have been prevalent in the results of different studies. We argue that these results represent some fundamental aspects of programmers' mental models.

\subsection{How Have Programmers’ Mental Models Been Studied?}
In this review, we analyze the state of the research field on programmers' mental models. We evaluate the coverage of research in terms of target systems and participant populations. Furthermore, we evaluate the study designs and the theoretical concepts used in the research.
\subsubsection{Participants, Experience and Target Systems}
We analyzed the target systems used in the studies. Our findings suggest the need to validate existing results using modern programming languages and environments. Our results show that many existing studies used programming languages such as BASIC and Fortran. Over the years, programming languages and tools have developed and new programming languages have risen to prominence.  However, we did not find studies using JavaScript, Go, or other similar programming languages that have recently gained popularity. Modern programming also relies heavily on different APIs and complex programming environments. These were almost completely missing from the target systems that have been studied.

Our results also show a focus on mental models of programs. Programmers' mental models of concepts and tasks have received less attention. Subsequent studies will have to further investigate programmers' mental models of programming tasks and concepts.

Our results show a good balance between experienced and novice participants being used. However, we categorized study participants according to the statements in the studies themselves. Therefore, we cannot guarantee the comparability of the classifications. We were surprised to see that only a few studies took into account task-specific experience. Most studies measured general computing experience (63 studies). Task-specific experience was measured only in 10 studies. These results call for further research to evaluate the role of task-specific experience in programmers' mental models and their acquisition.

The most common measures used to assess experience either measured length of experience in different ways or used self-reporting. Direct skills or knowledge assessments were rarely used. These measures can provide information on the amount of experience one has. However, they do not necessarily correlate with expertise in a meaningful way \cite{feigenspan2012measuring, baltes2018towards}. These results show a clear need to build a more informative way to assess experience.
The most common type of participant was student. Some of the studies in this set of results aimed to understand students' mental models, so the use of students as participants is expected. However, in some studies, students were used instead of or representing programmers in general. This practice has been debated~\cite{falessi2018empirical,el2019comparative}. However, it is also supported by parts of the research field~\cite{falessi2018empirical}.

\subsubsection{Study Designs}
Our results highlight the need to validate existing research results in natural, contemporary settings. Most of the included studies were carried out in laboratory settings. Data collected in laboratory settings is critical for research \cite{stol2015holistic}. However, research in natural settings is essential to confirm these results and their applicability to real-world situations. Modern software development settings use a wide range of tools and technologies. Older studies have not been validated in such settings. This points to the need to assess how modern software development settings affect programmers' mental models.

Non-coding tasks were the most prevalent study tasks. To assess the applicability of the results to real-world situations, the use of naturalistic programming tasks would be important. Researchers have made similar arguments about the field of software engineering research in general. They have called for realistic study designs to confirm the applicability of research results to real world settings \cite{sjoberg2002conducting}. 

\subsubsection{Theoretical Concepts}
 A large proportion of the studies did not rely on any theoretical concepts. We hypothesize that some of this is due to the lack of relevant theoretical concepts to refer to. In general, the use of theory in software engineering research is lacking \cite{hannay2007systematic, hall2009systematic}. This lack of background theory can cause problems in interpreting the study results, as no framework is used to explain and conceptualize the results \cite{detienne2001software}.

From the initial set of studies, we were able to identify 12 structure concepts and 15 acquisition concepts described in Table~\ref{tab:dataextractionform}. These concepts were not enough to cover the wide set of concepts and theories related to programmers' mental models. The ``other'' category was large for both theoretical concept categories even after adjustments were made during the snowballing phase. When analyzing the field of software engineering research, similar results have been achieved, showing little shared theory between studies \cite{hannay2007systematic}.

\subsection{How Have Programmers’ Mental Models Been Conceptualized and Described?}
The majority of the studies in our set of results were concerned with mental models of programs. This analysis also focused heavily on these mental models, as data related to mental models of other types of target system were limited.

Details of the mental models vary between studies. However, some core ideas remain constant. This shows a kernel of shared understanding of what mental models are. Mental models are understood as information or knowledge structures that represent programmers' knowledge or understanding of a program, concept, or programming situation. They are described as representing some important aspects of the target system. They are defined as either a collection of related knowledge or as a representation of the target system at multiple levels of abstraction. When it comes to mental models of programs, the types of information contained within them are typically described as the structure and functioning of the program. Information content related to mental models of other types of target systems is rarely described. The information content of task mental models was not described in this data set. We can conclude that \emph{programmers' mental models are considered to be meaningfully organized representations of programmers' understanding of aspects of a target system}.

The research field seems to agree on some of the essential functions of mental models. Mental models are seen as mental representations that are used when performing programming tasks such as debugging and modification. These mental models are described as tools that allow us to understand, interpret, and explain the target system through mental simulation and reasoning and that programmers use to make decisions and predictions. Thus, according to the research field \emph{mental models are used in performing tasks related to the target system: they allow explaining, predicting, and interpreting the target system, thus aiding in tasks such as making decisions and predictions}.

Mental model acquisition is, according to descriptions, not limited to program comprehension. They are acquired in various sense-making situations, such as educational settings, program comprehension, program design, and learning to use a new system. This process is described as iterative and active, using the programmer's background knowledge in some fashion. One key facet of the acquisition process seems to be mapping or linking features of the actual target system to its mental representation. This acquisition process and the resulting mental models were described as being informed by a combination of information present in the environment and the programmers' knowledge and experiences. In particular, the program code and other auxiliary materials, such as documentation, were mentioned as information sources.
Thus, \emph{mental model acquisition is understood to happen in various situations where the programmer makes sense of the target system. This process is done in an iterative fashion using the information available in the situation and the programmers' background knowledge and experience}.

This understanding of mental models is in line with how mental models are understood in other fields of research \cite{moon2019mental, van2021reflections}. Mental models have often been described as internal representations of some aspects of the world that are used in reasoning \cite{alfred2020mental} and guide behavior \cite{van2021reflections}. The spread of target systems is also present in the literature, with mental models described as representing both dynamic systems or situations, or more static structures or entities of the world \cite{van2021reflections}.

Although our research shows that the general understanding of programmers' mental models is somewhat unified and in line with research from other fields, the specifics of programmers' mental models and their use and acquisition in the context of programming remain undefined. Further research is required to evaluate the information content of programmers' mental models and the specifics of their acquisition to understand what is required to apply current knowledge of programmers' mental models in practical solutions. 

\subsection{Synthesizing existing knowledge on programmers' mental models}
With RQ3 and RQ4, ``What are the central aspects of programmers’ mental models present in the literature?'' and ``What can be concluded about programmers' mental models based on the results related to these central aspects?'' we synthesized current knowledge on programmers' mental models.

 Although vast amounts of data exist to support knowledge about some aspects of programmers' mental models, due to differences in terminology and theoretical backgrounds, so far it has been difficult to connect the results into a coherent picture. Through our analysis, we were able to identify eight aspects of programmers' mental models that are highlighted in research results across the research field and which represent some fundamental aspects of programmers' mental models.

On the basis of our results, we can describe programmers' mental models as mental representations of knowledge about the target system. When it comes to mental models of programs, we see the need for both syntactic and semantic information about the program, although the prevalence of either type of information depends on the programmer, their level of expertise, and the specific sense-making situation.

In line with the descriptions of mental models in the literature, mental models are acquired utilizing information present in the environment and the programmers' background knowledge. This background knowledge contains information on programming concepts, patterns, and syntax and is developed through education and practice. The lack of this background knowledge is one of the factors affecting novice programmers' performance. They cannot detect familiar programming patterns from code and utilize them in programming tasks.

Program comprehension is an active process of seeking information according to the programmer's goals and information needs from multiple information sources. Programmers have been shown to perform a variety of actions, from reading, scanning, and searching the codebase, to mental simulation, and reasoning to gather this information.
The program comprehension process is driven by hypotheses. These hypotheses are generated based on background knowledge and features of the target system, and represent many aspects of the target system from the target system structure to the location and relevance of different parts of the system.

Some features of the code act as indicators that are used to detect specific features of the code. Indicators come in many forms, from specific lines of well-known code patterns to file names, comments, and function and variable names. These indicators are used to form expectations about the code and verify hypotheses by top-down recognition of code elements.

When programming or designing programs, programmers construct mental representations of the code they are creating. These mental models start from a kernel idea that contains the key facets of the program and are then systematically extended to a more complete representation.

We argue that while the existing theories on programmers' mental models present a somewhat accurate picture of programmers' mental models, these theories need to be reevaluated when it comes to the importance of locating code and the information needs of programmers in real-world programming contexts. In terms of programmers' mental models of programs, these results most correspond with Von Mayrhauser's integrated metamodel theory of programmers' mental models \cite{von1995industrial}.  However, the model lacks some aspects that we uncovered in our analysis.

In our analysis, we see that the location of specific code elements is an important information need.  Programmers search and scan code to find specific code elements. They also create hypotheses about their location based on indicators and their background knowledge \cite{latoza2007program, jeffries1982comparison, burgos2007through, sillito2008asking}. This aspect of locating code elements is missing from integrated metamodel theory. Some of the other information needs that were present in our analysis are also not present in the integrated metamodel theory. In particular, the need of programmers to understand the rationale of the program and how the program has changed and is changing over time \cite{burgos2007through, yates2020characterizing, levy2019understanding} is not found in the integrated metamodel theory.

\subsection{Threats to validity}
\label{S:6}
\input{57-validity}

%% file: 57-validity.tex
In this section, we discuss threats to the validity of our research and describe the ways in which we mitigated these threats. 

The design of our search string could lead to omitting some relevant articles. It needed to catch relevant articles but simultaneously omit irrelevant results. We opted to allow the search string to be broad and relied on our exclusion criteria to remove irrelevant results.
We also complemented the database searches with backward and forward snowballing as discussed in Sections~\ref{sec:database-search} and~\ref{sec:snowballing}.

Our article selection and filtering process is a source of bias. During database searches, a researcher conducted the first exclusion round. Care was taken to exclude only obvious false positives, such as results not written in English and results that were not the required publication type, such as posters or abstracts. In the other phases of the filtering process, we followed well-established SLR guidelines~\cite{keele2007guidelines}. As discussed in Section~\ref{sec:inclusion-exclusion-criteria}, the filtering was done by a team of multiple researchers and a verification step was used to mitigate the possible biases of any single researcher. 

The data extraction process was another source of bias. To ensure consistency in data extraction,  we first ran a round of practice extraction sessions. In these sessions, each researcher processed the same articles individually. The resulting extractions were discussed with the other researchers to reach a consensus on how to interpret the extraction guidelines. During database searches, each researcher extracted data from articles in batches of six.  Another researcher verified one extracted article from each batch, as well as all excluded articles. During the snowballing phase, the batch size was increased to 10 as the team became more familiar with the process. Whenever a discrepancy was detected, the interpretation was discussed between the extractor and the verifier or, if uncertainties remained, with the entire team. If it was discovered that the discrepancy affected more than one paper, the extractor would return to the affected papers to adjust the extraction accordingly, and the verification was repeated. The emphasis was on detecting and repairing inconsistencies in extraction, not measuring extraction error. Therefore, we did not calculate interrater agreement between extractors.

Our data extraction includes data that was summarized from the articles by each extractor, such as the mental model description data. These extractions can be subjective, biasing the results. We used thematic analysis~\cite{guest2012} to analyze descriptions of mental models and summaries of results related to mental models. Two researchers participated in the analysis to mitigate some risks posed by the subjectivity of the researchers. The results of the analysis were presented to the other researchers for validation. Given the extraction verification process and the way the data was analyzed, we argue that the reliability of the data is adequate for our purposes.

%% file: 60-conclusions.tex
This systematic literature review assessed the state of the research field on programmers' mental models to bring consistency into the fragmented knowledge base related to them. Our results show that to achieve an understanding of programmers' mental models that applies to real-world settings, more research is required. Up to this point, research has focused on programmers' mental models of programs and relied on studies using non-programming tasks in laboratory settings. These studies have been conducted using programming languages and environments that have fallen out of fashion.  So far, the effects of modern languages and tools have received little attention.

Although the importance of researching programmers' mental models to understand and support their work has been recognized by the research community, programmers' mental models of code, tasks, and programming concepts have so far been studied from mutually inconsistent starting points, producing a largely scattered set of findings. This fragmentation in the findings and the lack of a coherent theoretical basis cause difficulties in advancing the research field into analyzing and interpreting comprehension and the generation of large-scale software in modern settings.

In this work, we have untangled some of the fragmentation in the field. Our analysis shows that across the research field, mental models are understood in terms of meaningfully organized representations of programmers' knowledge of some aspects of a target system, which are used in performing tasks related to the target system. We have also detailed fundamental aspects of mental models that are prominent in results from across the research field. We argue that these aspects can be used to build a shared understanding of mental models to use as a basis for further research. 

\section*{Declaration of Competing Interest}
The authors declare that they have no known competing financial interests or personal relationships that could have appeared to influence the work reported in this article.

\section*{Credit Author Statement}
\textbf{Ava Heinonen:} Conceptualization, Methodology, Validation, Formal analysis, Investigation, Data Curation, Writing - Original Draft, Writing - Review \& Editing, Visualization, Supervision.
\textbf{Bettina Lehtelä:} Methodology, Validation, Formal Analysis, Investigation, Data Curation, Writing - Original Draft, Writing - Review \& Editing.
\textbf{Arto Hellas:} Validation, Formal analysis, Investigation, Data Curation, Writing - Original Draft, Writing - Review \& Editing, Visualization.
\textbf{Fabian Fagerholm:} Methodology,  Validation, Investigation, Data Curation, Writing - Original Draft, Writing - Review \& Editing, Supervision.

\section*{Acknowledgments}
We thank Bal\'{a}zs Nagyv\'{a}radi, Jehan Khattak and Markus Laattala for their help in initial study filtering and data extraction.